\documentclass[12pt]{article}
\usepackage{epsfig}
\title{SO(8) Colour as possible origin of generations}
\author{Z.~K.~Silagadze \vspace*{2mm}\\
\small\rm Budker Institute of Nuclear Physics,
  630 090, Novosibirsk, Russia}
\date{}

\begin{document}
\large\rm
\maketitle

\begin{abstract}
 A possible connection between the existence of three quark-lepton
 generations  and  the  triality property  of  SO(8)  group  (the  equality
 between 8-dimensional vectors and spinors) is investigated.
\end{abstract}

\section{Introduction}
  One  of  the  most  striking  features  of   quark-lepton spectrum is its
 cloning property: $\mu$ and $\tau$ families seem  to be just heavy copies
 of electron family. Actually  we  have two questions to be answered: what
 is an origin  of  family formation and how many generations  do  exist.
  Recent  LEP  data \cite{1} strongly suggests three quark-lepton generations.
 Although Calabi-Yau  compactifications  of  the  heterotic string model can
 lead to three generations \cite{2}, there are many such Calabi-Yau manifolds,
 and additional  assumptions are needed to argue why the number three is
 preferred \cite{3}.

  There is another well-known example of  particle  cloning (doubling  of
  states):  the  existence  of  antiparticles. Algebraically the charge
 conjugation  operator  defines  an (outer) automorphism of underlying
  symmetry  group \cite{4,5} and reflects the symmetry of the corresponding
 Dynkin  diagram. We can thought that the observed triplication of states
 can have the same origin.

  The most symmetric  Dynkin  diagram  is  associated  with SO(8) group.
 So it is the richest in automorphisms  and  if SO(8) plays some  dynamical
  role  we  can  hope  that  its greatly symmetrical internal structure
  naturally  lead  to the desired multiplication of states in elementary
 particle spectrum. what follows is an elaboration of this idea.

  Although the relevant mathematical  properties  of  SO(8) are known for
 a long time \cite{6}, they have not  been  discussed in context of the
 generation problem, to my knowledge.

\section{Peculiarities of the SO(8) group}
  It is well known \cite{7,8} that the structure of  a  simple  Lie algebra
 is  uniquely  defined  by  the  length  and  angle relations among simple
 roots. This information is compactly represented by the Dynkin diagram.
 On such a  diagram  each simple root is depicted by a small circle,
  which  is  made black, if the root is a short one. Each pair of vertexes
 on the Dynkin diagram is connected by lines, number  of  which equals to
 $4\cos^2{\varphi}$, $\varphi$ being the angle  between corresponding simple
 roots.

  The main classification theorem for simple  Lie  algebras states that there
 exists only four infinite series and five exceptional algebras \cite{7}.
 Among them $D_4$, the Lie algebra of the SO(8) group, really
 has the most symmetric Dynkin diagram 
\begin{figure}[htb]
     \centerline{\epsfxsize 60mm\epsfbox{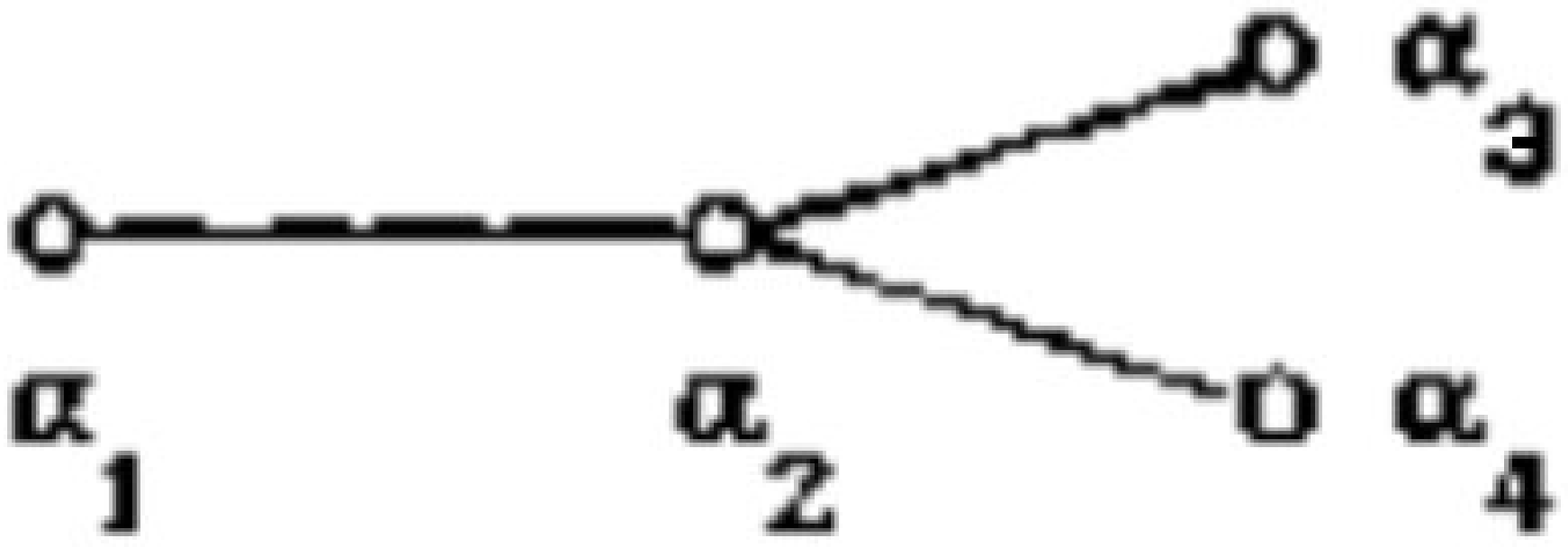}}
\end{figure}

  Actually only the symmetry with regard  to  the  cyclic permutations of
the $(\alpha_1 ,\alpha_3 ,$ $\alpha_4)$ simple roots (which we call triality
 symmetry) is new, because the symmetry with regard to the interchange
  $\alpha_3 \longleftrightarrow \alpha_4$ (last two  simple  roots)  is
 shared by other $D_n$  Lie algebras also.

  Due to this triality symmetry, $8_v =(1000),8_c =(0010)$  and
 $8_s =(0001)$ basic irreps ( $(a_1 ,a_2 ,...,a_r)$  being  the  highest
 weight in the Dynkin  coordinates \cite{8}) all have the same dimensionality
 8 - the remarkable fact valid only  for  the $D_4$  Lie  algebra.
  The  corresponding  highest  weights  are connected by above mentioned
 triality symmetry.  For  other orthogonal groups  (10\ldots0)  is  a  vector
 representation, (00\ldots01) - a first kind spinor and (00\ldots10)  -  
 a  second kind spinor. So there is no  intrinsic  difference  between 
 (complex) vectors and spinors in the eight-dimensional space \cite{9}, which 
 object is vector and which ones are  spinors depends simply on how we have
 enumerated  symmetric  simple roots and so is a mere convention.

  It is tempting to use this peculiarity of the SO(8) group to justify
 observed triplication of quark-lepton degrees of freedom. This possible
 connection between  generations  and SO(8)  can  be  formulated  most
 naturally  in  terms   of octonions.

\section{Octonions and triality}
  Eight-dimensional vectors and  spinors  can  be  realized through
octonions \cite{10,11} , which  can  be   viewed   as   a generalization
 of  the  complex  numbers:  instead  of  one imaginary unit we have seven
 imaginary units $e_{A}^{2} =-1,\hspace*{2mm}  A=1 \div 7$, in the octonionic
 algebra. The multiplication table between them can be found in \cite{11}.

  The octonion algebra is an alternative algebra  (but  not associative).
    This    means    that    the    associator
 $(x,y,z)=x \cdot (y \cdot z)-(x \cdot y) \cdot z$ is a skew symmetric
 function of the x,y,z octonions.

  the conjugate  octonion  $\bar q$  and  the  scalar  product  of octonions
 are defined as
\begin{eqnarray}
   \bar q=q_0 -q_A e_A  \hspace*{5mm} (p,q)=\frac{1}{2}(p\cdot \bar q+
q\cdot \bar p)=(\bar p, \bar q)
 \label{eq1}\end{eqnarray}

  Let us consider  eight  linear  operators   $\Gamma_m $, \hskip2pt
$ m=0 \div 7 $, acting in the 16-dimensional bioctonionic space:
\begin{eqnarray}
\Gamma_m \pmatrix{q_1 \cr q_2 \cr}=\pmatrix{0 & e_m \cr \bar e_m & 0 \cr}
\pmatrix{q_1 \cr q_2 \cr}=\pmatrix{e_m \cdot q_2 \cr \bar e_m \cdot q_1 \cr}
\label{eq2}\end{eqnarray}

  Using the alternativity property of octonions,  it  can  be tested that
 these operators generate a Clifford algebra
  $$ \Gamma_m \Gamma_n + \Gamma_n \Gamma_m = 2 \delta_{mn} \hskip5pt . $$
(Note, that,  because  of  nonassociativity,  the  operator product  is  not
  equivalent  to   the  product  of  the corresponding octonionic matrices).

  The  eight-dimensional  vectors  and   spinors   can   be constructed  in
 the  standard  way from  this  Clifford algebra \cite{12}.  Namely,  the
  infinitesimal   rotation   in   the (k,l)-plane by an angle $\theta$ is
 represented by the operator
 $$ R_{kl}=1+\frac{1}{2}\,\theta\, \Gamma_k \Gamma_l \hskip5pt , $$
                            and the transformation law for  the  (bi)spinor
 $\Psi=\pmatrix{q_1 \cr q_2 \cr}$  is $\Psi ^{\prime}=R_{kl}\Psi $.

  For $\Gamma_m$ given by  Eq.2  the  upper  and  lower  octonionic
 components  of $\Psi$ transform   independently   under   the 8-dimensional
 rotations:
\begin{eqnarray}
&&q^{\prime}_1=q_1+\frac{1}{2}\,\theta \, e_k\cdot(\bar e_l \cdot q_1)
\equiv q_1+\theta F_{kl}(q_1) \nonumber \\[2mm]
&&q^{\prime}_2=q_2+\frac{1}{2}\,\theta \, \bar e_k\cdot(e_l \cdot q_2)
\equiv q_1+\theta C_{kl}(q_2)
 \label{eq3}\end{eqnarray}
while a vector transformation law can be represented in the form
\begin{eqnarray}
x^{\prime}=x+\theta \,\lbrace e_k(e_l,x)-e_l(e_k,x) \rbrace
\equiv x+G_{kl}(x)
\label{eq4} \end{eqnarray}

   One  more  manifestation   of   the   equality   between 8-dimensional
 vectors and spinors is the fact \cite{9} that each spinor transformation
 from Eq.3 can be represented as a sum of four vector rotations
\begin{eqnarray}
 F_{0A}=\frac{1}{2}(G_{0A}+G_{A_1B_1}+G_{A_2B_2}+G_{A_3B_3})
\label{eq5}\end{eqnarray}
where $A_i ,B_i$  are defined through the condition 
$e_{A_i}\cdot e_{B_i}=e_A$, and
\begin{eqnarray}
 F_{A_1B_1}=\frac{1}{2}(G_{A_1B_1}+G_{0A}-G_{A_2B_2}-G_{A_3B_3})
\label{eq6}\end{eqnarray}

  An algebraic expression of the equality  between  vectors and spinors in
 the eight-dimensional space is the following equation, valid for any two
 $x,\;y$ octonions \cite{11}:
\begin{eqnarray}
\overline {\rm S_{kl}(\overline {\rm x\cdot y})}=G_{kl}(x)\cdot y+x\cdot
C_{kl}(y)
 \label{eq7}\end{eqnarray}
where  $S_{kl}=KF_{kl}K$, \hskip2pt K being  the  (octonionic) conjugation
 operator $K(q)=\bar q$.

  Eq.7 remains  valid  under  any  cyclic permutations of
 $(S_{kl},G_{kl},C_{kl})$. Note that
\begin{eqnarray}
 S_{kl}=\tau (G_{kl}), \;\; C_{kl}=\tau (S_{kl})=\tau ^2(G_{kl})
\label{eq8}\end{eqnarray}
where $\tau$ is an automorphism of the $D_4$  Lie  algebra.  We  can call it
 the triality automorphism, because  it  performs  a cyclic  interchange
  between  vector   and   spinors:   $G_{kl}$   operators realize
 the (1000) vector representation, $S_{kl}$ - a first kind spinor (0001),
 and $C_{kl}$   - a  second  kind  spinor (0010).

  In general,  vector  and  spinors  transform  differently under
 8-dimensional rotations, because $G_{kl}\not=S_{kl}\not=C_{kl}$ . But  it
 follows from Eq.6  that  $G_{A_1B_1}-G_{A_2B_2}$ and
  $G_{A_1B_1}-G_{A_3B_3}$    are invariant with regard to the triality
  automorphism, and  so under such rotations 8-dimensional vector and both
 kinds of spinors transform in the same  way.  These  transformations are
 automorphisms of the octonion  algebra,  because  their generators act as
 derivations, as the principle of triality (Eq.7) shows. We  can  construct
  14  linearly  independent derivations of the octonion  algebra,  because
  the  method described above gives two  independent  rotations  per  one
 imaginary octonionic unit $e_A=e_{A_i}\cdot e_{B_i}$. It  is  well known
 \cite{10} that the  derivations  of  the  octonion  algebra  form  $G_2$
  exceptional Lie algebra. It was suggested \cite{11,13,14} that the subgroup
 of this $G_2$ ,which leaves the seventh imaginary unit invariant, can be
 identified with the colour  SU(3)  group. If we define the split octonionic
 units \cite{11}   \begin{eqnarray}
&&u_0=\frac{1}{2}(e_0+ie_7) \hskip24mm u_{0}^*=\frac{1}{2}(e_0-ie_7)
\nonumber\\[2mm]
&&u_k=\frac{1}{2}(e_k+ie_{k+3}) \hskip20mm u_{k}^*=\frac{1}{2}(e_k-ie_{k+3})
\label{eq9}\end{eqnarray}
where $k=1\div3$, then with regard to this SU(3)  $u_k $  transforms as
 triplet, $u_{k}^*$  - as antitriplet and $u_0$ ,$u_{0}^*$ are singlets
 \cite{11}. Therefore, all one-flavour quark-lepton degrees of  freedom can
 be represented as one octonionic (super)field
\begin{eqnarray}
     q(x) = l(x)u_0 +q_k(x)u_k +q_{k}^C(x)u_{k}^*+l^C(x)u_{0}^*
\label{eq10}\end{eqnarray}
here $l(x),\;q_k(x)$  are  lepton  and  (three  coloured)  quark fields and
 $l^C(x), \;q_{k}^C$  - their charge conjugates.

  Note that it doesn't matter what an octonion, first  kind spinor, second
 kind spinor or  vector  we  have  in  Eq.10, because they all transform
 identically under SU(3).

  So SO(8) can  be  considered  as  a  natural $\rm \underline{one-flavour}$
 quark-lepton unification group.  We  can  call  it  also  a generalized
 colour  group   in   the   Pati-Salam   sense, remembering their idea  about
  the  lepton  number  as  the fourth colour \cite{15}. Then the triality
 property  of  the  SO(8) gives a natural reason why the number of flavours
 should be triplicated.

\section{Family formation and SO(10)}
  Unfortunately, SO(8) is not large enough to be used as  a grand
  unification  group:  there  is  no  room  for   weak interactions in it.
 This is not  surprising,  because  weak interactions connect two  different
  flavours  and  we  are considering SO(8) as a one-flavour unification group.

  The following observation points out the  way  how  SO(8) can be extended to
 include the weak  interactions.  Because $C_{AB}=F_{AB}$ and
 $C_{A0}=-F_{A0}$   for  $A,B=1 \div 7$,  the  SO(8)  (Hermitian) generators
 for the (bi)spinor transformation  Eq.3  can  be represented as
 $M_{AB}=-iF_{AB}$ and $M_{A0}=-M_{0A}=-i \sigma_3 F_{A0}$.  The  last
 equation suggests  to  consider  $M_{A,7+k} =-i \sigma_k F_{A0}$
   generators, where $k=1 \div 3$ and summation to the  modulus  10  is
  assumed, i.e. 7+3=0. So  we  have  two  new  operators
  $-i \sigma_1 F_{A0}$ and $-i \sigma_2 F_{A0}$ which mix the upper and lower
 (bi)spinor octonionic components. Besides, if  we  consider these  operators
  as rotations, then we have to add two extra dimensions and  it is expected
 that SO(8) will be enlarged to SO(10)  in  this way  and  two  different
  SO(8)  spinors   (two   different flavours)  will  join  in   one   SO(10)
   spinor   (family formation).

  Indeed, the following generators
\begin{eqnarray}
&& M_{AB} =-iF_{AB}   \hskip10mm   M_{7+i,7+j}=\frac{1}{2} \varepsilon_
{ijk} \sigma_k \nonumber \\[2mm]
&& M_{A,7+k}=-i \sigma_k F_{A0} \hskip5mm M_{7+k,A}=-M_{A,7+k}
\hskip5pt , \label{eq11} \end{eqnarray}
where $A,B=1 \div 7$ and $i,j,k=1 \div 3$ ,  really  satisfy  the  SO(10)
 commutation relations
\begin{eqnarray}
 [M_{\mu \nu},M_{\tau \rho}]=-i(\delta_{\nu \tau} M_{\mu \rho}
 +\delta_{\mu \rho} M_{\nu \tau} - \delta_{\mu \tau}
  M_{\nu \rho} - \delta_{\nu \rho} M_{\mu \tau }) \hskip3pt  .
\label{eq12} \end{eqnarray}

  It is clear from Eq.12, that $M_{\alpha \beta}$ $(\alpha , \beta=0,7,8,9)$
 and  $M_{mn}$ $(m,n=1 \div 6)$  subsets   of   generators   are   closed
  under commutation and commute to each other. They  correspond  to
$SU_L(2) \otimes SU_R(2)$ and SU(4) subgroups of SO(10). The generators of
 the $SU_L(2) \otimes SU_R(2)$ can be represented as
\begin{eqnarray}
T_{L}^i =\frac{1}{2} \sigma_i u_0 \hskip10pt , \hskip10pt T_{R}^i=\frac{1}{2}
\sigma_i u_{0}^*
\hskip5pt  . \label{eq13}  \end{eqnarray}
So multiplication by $u_0$ or $u_{0}^*$  split  octonion  units  plays the
 role of projection operator on the left and right  weak isospin,
 respectively.

  The SU(4) generators can  be  also  expressed  via  split octonionic units:
\begin{eqnarray}
  E_{ij} =-u_i \cdot (u_{j}^* \hskip20pt , \hskip20pt E_{0i}=-u_j \cdot (u_k
\hskip20pt , \hskip20pt E_{i0}=u_{j}^*
\cdot (u_{k}^* \hskip40pt .
\label{eq14} \end{eqnarray}
In the last two equations $(i,j,k)$ is a  cyclic  permutation of  (1,2,3)
 and  it  is   assumed   that,   for   example,
 $E_{ij}(q)=-u_i \cdot (u_{j}^* \cdot q)$ .

  Under SU(4) $u_ \alpha$  , $\alpha=0 \div 3$, transforms as  a
$\rm \underline 4 $ fundamental representation and $u_{\alpha}^*$ - as  its
  conjugate  $\rm \underline{4^*}$ .  So  SU(4) unifies $u_0$  colour singlet
 and  $u_k$   colour  triplet  in  one single  object,  and  therefore  plays
 the  role  of   the Pati-Salam group \cite{15}.

  Note  that  all one-family  (left-handed)   quark-lepton degrees  of
  freedom  are  unified  in   one   bioctonionic (super)field (16-dimensional
 SO(10) spinor) \cite{16}
\begin{eqnarray}
\Psi_L=\pmatrix{\nu (x) \cr l(x) \cr}_L u_0 +
\pmatrix{q_{i}^u (x) \cr q_{i}^d(x) \cr}_L u_i +
\pmatrix{l^C (x) \cr \nu^C (x) \cr}_L u_{0}^* +
\pmatrix{q_{i}^{dC}(x) \cr q_{i}^{uC}(x) \cr}_L u_{i}^*
\hskip2pt . \label{eq15} \end{eqnarray}

The fact that we should take the Weyl (left-handed) spinors instead of Dirac
 (that the weak  interactions  are  flavour chiral) indicates close interplay
 between space-time (space inversion) and internal symmetries \cite{17}.

  Thus our  construction  leads  to  SO(10)  as  a  natural
$\rm \underline {one-family}$ unification group. But doing so, we have
  broken the triality symmetry: only  the  spinoric  octonions  take part in
 family  formation  and  the  vectoric  octonion  is singled out. Can we in
 some way restore equivalence between vector and spinor octonions?

  First of all we need to realize vector octonion in  terms of the SO(10)
 representation and this can be done by  means of $2 \times 2$ octonionic
 Hermitian matrices, which  together  with the symmetric  product
 $X \circ Y=\frac{1}{2}(XY+YX)$ form the $M_{2}^8$ Jordan algebra \cite{18}.
 SO(10) appears as a (reduced) structure group of this Jordan algebra
 \cite{18} and  10-dimensional  complex  vector space   generated  by the
  $M_{2}^8$  basic elements   (the complexification  of   $M_{2}^8$ ),
 gives  (10000)  irreducible representation of its $D_5$  Lie algebra.

  Thus, now we have at hand  the  realization  of  spinoric octonions  as  a
   16-dimensional  SO(10)  spinor $\pmatrix{q_1 \cr q_2 \cr}$ and vectoric
 octonion as a 10-dimensional SO(10) vector
 $\left[ \matrix {\alpha & q \cr \bar q & \beta \cr } \right]$. How to unify
 them? The familiar  unitary  symmetry  example how to unify an isodublet and
  an  isotriplet  in  the  $3\times 3$ complex Hermitian matrix can give a
 hint  and  so  let  us consider $3\times $3 octonion Hermitian matrices.

\section{$E_6$, triality and family triplication}
  Together  with  the  symmetric  product,   $3\times 3$  octonion Hermitian
  matrices  form   the   $M_{3}^8$ exceptional  Jordan algebra \cite{10}.
 A  general  element  from  it has a form
$$X=\pmatrix{\alpha & x_3 & \bar {x}_2 \cr \bar {x}_3 & \beta & x_1
\cr x_2 & \bar {x}_1 & \gamma \cr}$$
 and can  be  uniquely represented as
 $X=\alpha E_1+ \beta E_2+ \gamma E_3+F_{1}^{x_1}+F_{2}^{x_2}+F_{3}^{x_3}$.
This is the Peirce decomposition \cite{19} of $M_{3}^8$   relative  to  the
mutually orthogonal idempotents $E_i$ .

  A reduced structure group of $M_{3}^8$ is  $E_6$   exceptional  Lie group
 \cite{18}.  Its  Lie  algebra  consists  of the following transformations:

\noindent
1) 24 linearly independent $\lbrace a_1,a_2,a_3 \rbrace$ generators, which
 are defined as \linebreak $\lbrace a_1,a_2,a_3 \rbrace X=[A,X]$, where
$$A=\pmatrix{0 & a_3 & \bar{a}_2 \cr - \bar{a}_3 & 0 & a_1 \cr
- \bar{a}_2 & - \bar{a}_1 & 0 \cr }$$
is a $3\times 3$ octonion anti-Hermitian matrix with zero  diagonal
elements.

\noindent
2)  $\lbrace \Delta_1,\Delta_2,\Delta_3 \rbrace$  triality  triplets
 (Eq.7  and  Eq.8),which annihilate $E_i$  idempotents and in the $F_i$
   Peirce  components act according to
$$\lbrace \Delta_1,\Delta_2,\Delta_3 \rbrace F_{i}^a=
F_{i}^{\Delta_i(a)}  .$$
Because a triality triplet is uniquely defined by its first
element:  $\Delta_2 =\tau(\Delta_1)$ and $\Delta_3=\tau(\Delta_2)$,
  this  gives  extra   28 linearly independent generators. Together  with
 $\lbrace a_1,a_2,a_3 \rbrace$ type operators, they form 52-dimensional
 $F_4$  exceptional Lie algebra \cite{10,20}.

\noindent
3) $T^{\wedge}$  linear transformations of $M_{3}^8$, defined  as
 $T^{\wedge}X=T \circ X $, where T is any element from $M_{3}^8$  with zero
 trace.

  The way how $E_6$  exceptional Lie  algebra  was  constructed shows the
 close relationship between $D_5$ and $E_6$: the latter is connected to the
 exceptional Jordan algebra $M_{3}^8$   and  the former - to the  $M_{2}^8$
 Jordan algebra \cite{21} . But $M_{3}^8$  has  three  $M_{2}^8$   (Jordan)
 subalgebras, consisting   correspondingly    from elements
$$\pmatrix{\alpha & a & 0 \cr \bar a & \beta & 0 \cr 0 & 0 & 0 \cr}
, \hskip5mm
\pmatrix{\alpha & 0 & \bar a \cr 0 & 0 & 0 \cr a & 0 & \beta \cr}
\hskip2mm and \hskip2mm
\pmatrix{0 & 0 & 0 \cr 0 & \alpha & a \cr 0 & \bar a & \beta \cr},$$
therefore $E_6$  has three equivalent $D_5$  subalgebras. Let $D_{5}^i$  be
that $D_5$  subalgebra of $E_6$   which  acts  in  the $M_{2}^8$   Jordan
algebra, formed from the $F_{i}^a,\,E_j,\,E_k$   elements.  It  consists
from  $\lbrace \Delta_1 ,\Delta_2 ,\Delta_3 \rbrace$, $(F_{i}^a)^ \wedge$
 ,$(E_j-E_k)^ \wedge$, $\lbrace \delta_{i1}a_1,\delta_{i2}a_2,\delta_{i3}a_3
\rbrace$ operators  and   their   (complex)   linear   combinations. Therefore
 the intersection of these $D_{5}^i$ subalgebras is $D_4$  formed from the
 $\lbrace \Delta_1,\Delta_2,\Delta_3 \rbrace$  triality  triplets,  and
  their unification gives the whole $E_6$ algebra.

  The triality automorphism for $D_4$  can be continued on $E_6$:
\begin{figure}[htb]
     \centerline{\epsfxsize 150mm\epsfbox{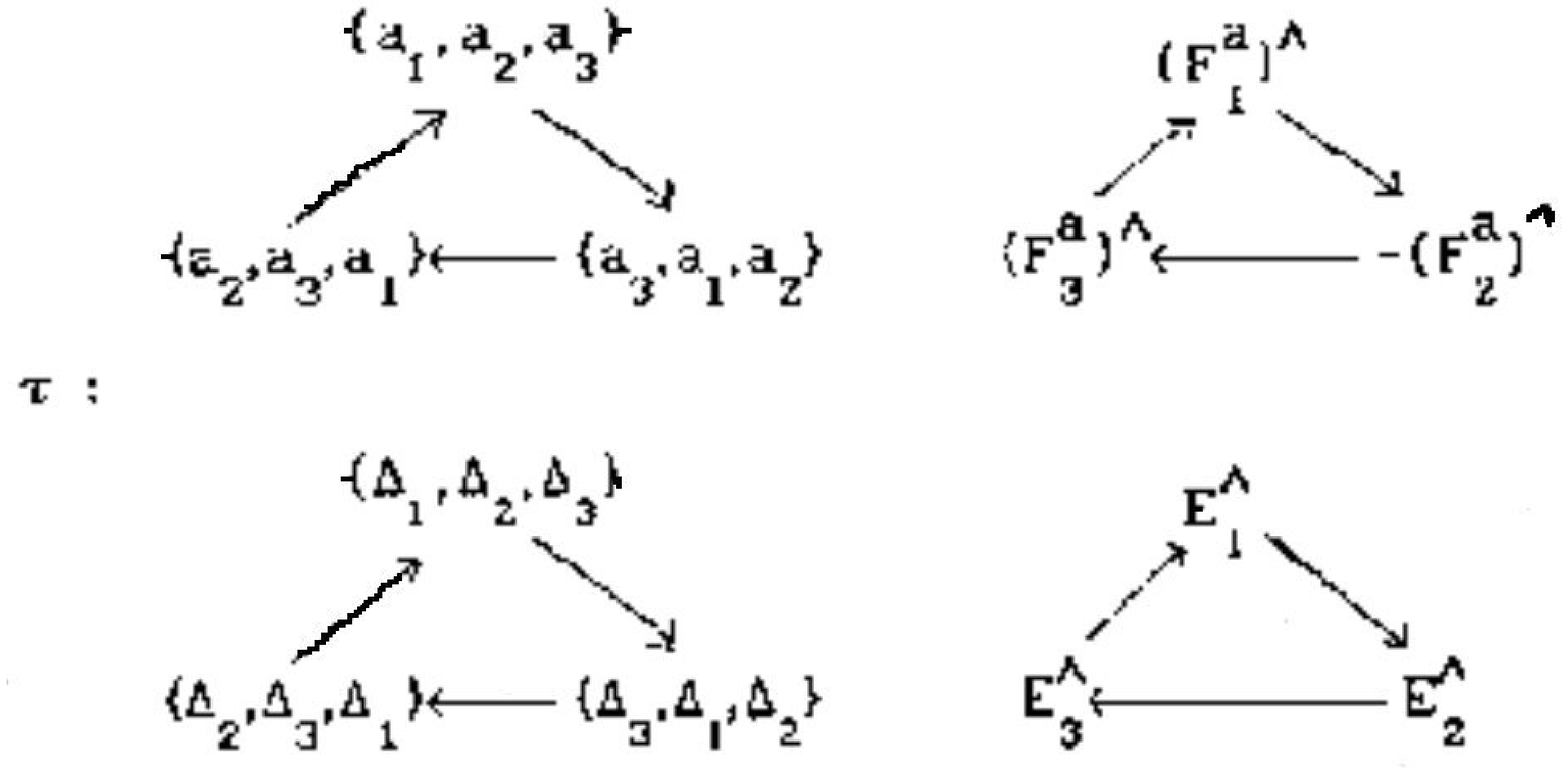}}
\end{figure}

It can be verified \cite{22} that Eq.16 actually gives an $E_6$ automorphism.
This  $\tau$   automorphism   causes   a   cyclic permutation of the $D_{5}^i$
subalgebras 
\begin{figure}[htb]
     \centerline{\epsfxsize 80mm\epsfbox{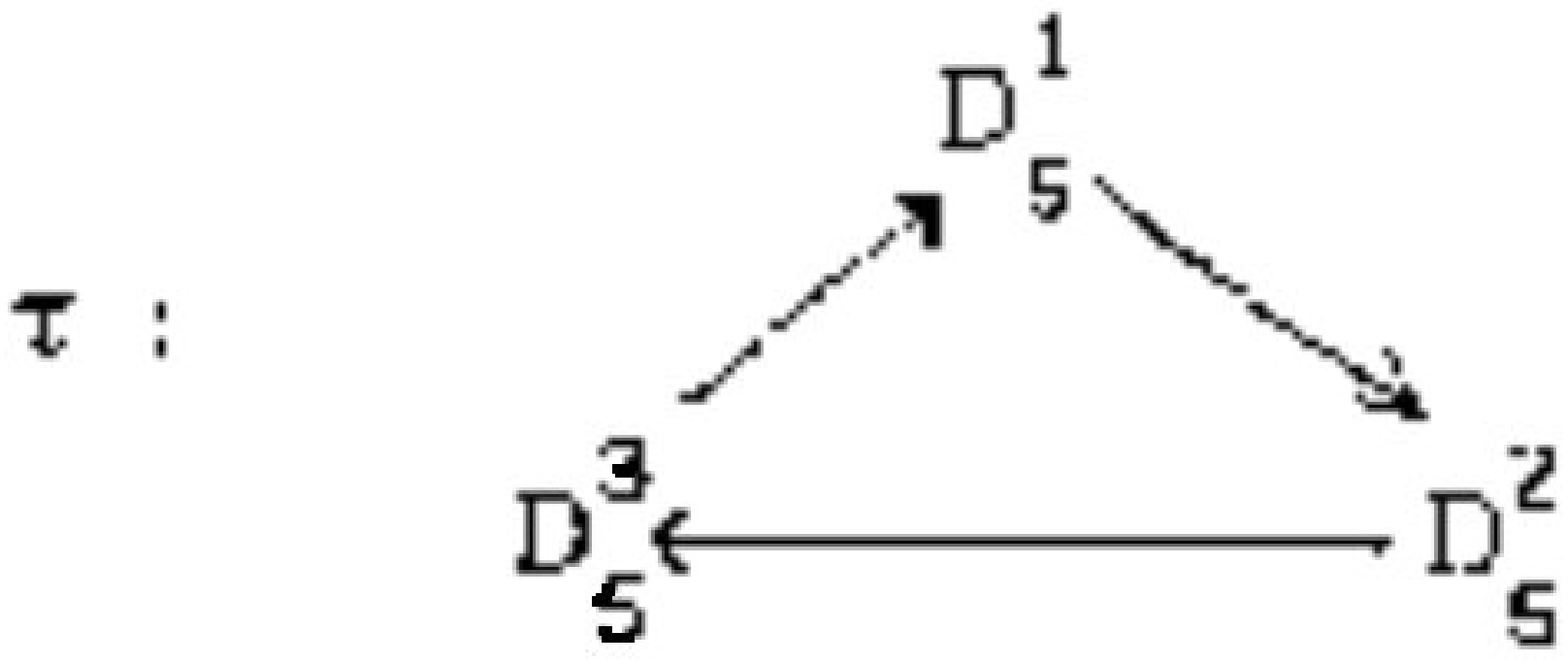}}
\end{figure}

   So $E_6$  exceptional Lie group is very  closely  related  to triality.
 Firstly, it unifies  the  spinoric  and  vectoric octonions in one
 27-dimensional irreducible representation (algebraically they unify in the
 $M_{3}^8$ exceptional  Jordan algebra). Secondly, its internal structure
 also  reveals  a very interesting triality picture:
\begin{figure}[htb]
     \centerline{\epsfxsize 60mm\epsfbox{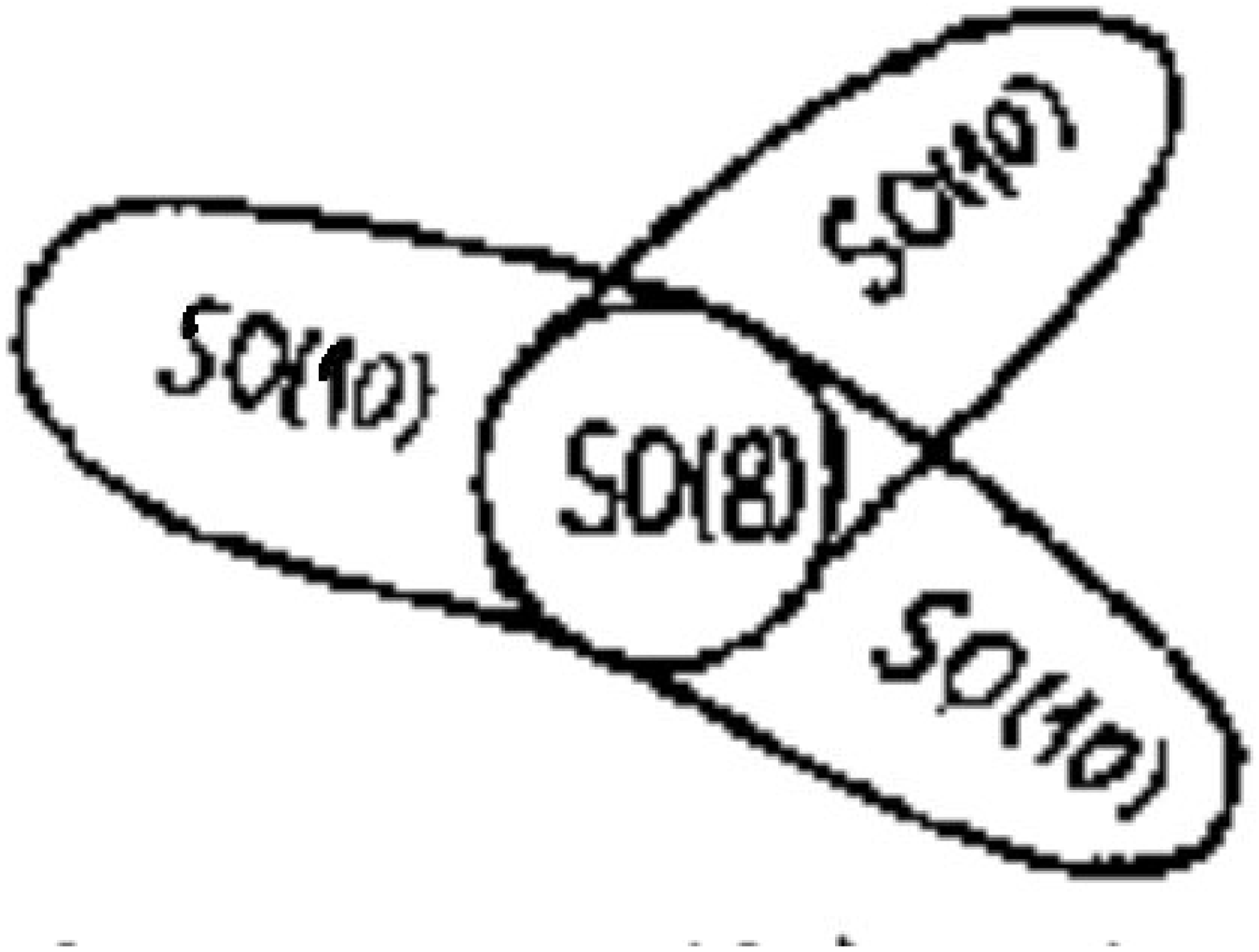}}
\end{figure}

 The  equality  between  SO(8)  spinors  and  vectors  now results in the
 equality of three SO(10) subgroups (in  the existence   of  the  triality
 automorphism  $\tau$,   which interchanges these subgroups).

  To form a quark-lepton family, we have to select  one  of these SO(10)
 subgroups. But a priori there is no reason  to prefer any of them. The
 simplest possibility to have family formation which  respects  this
  equality  between  various SO(10) subgroups ($E_6$ triality symmetry) is
  to  take  three copies of $M_{3}^8$ and arrange matters in such a way that
 in  the first $M_{3}^8$ the  first  SO(10)  subgroup  acts  as  a  family
 formating group, in the second $M_{3}^8$ - the second SO(10) and in the
 third one  -  the third SO(10):
\begin{figure}[htb]
     \centerline{\epsfxsize 120mm\epsfbox{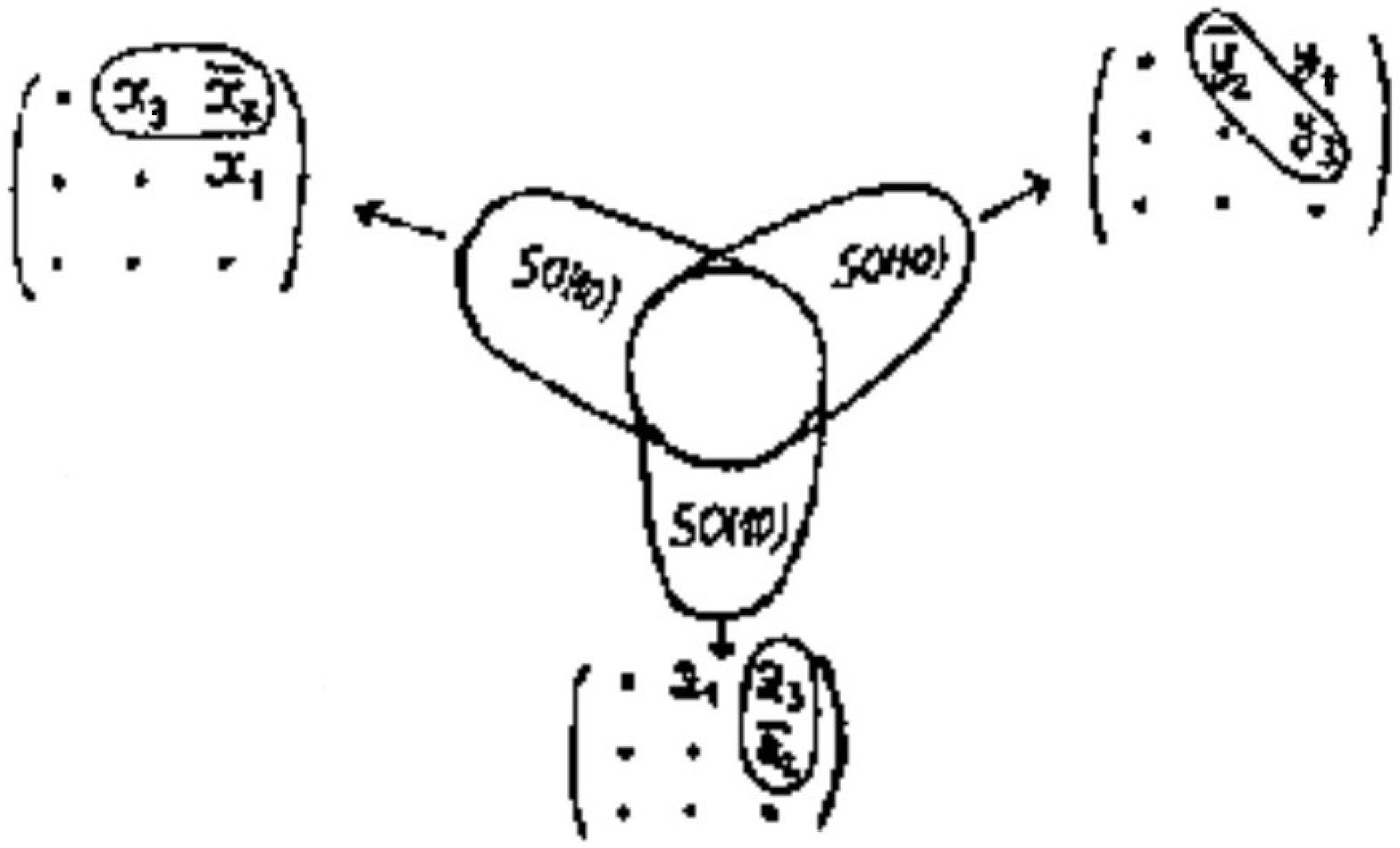}}
\end{figure}

  More formally, we have $\rm \underline{27}$+$\rm \underline{27}$+
$\rm \underline{27}$ reducible  representation of $E_6$ , such that when we
 go from one irreducible  subspace to another, representation  matrices  are
  rotated  by  the triality automorphism $\tau$.

\section{Conclusion}
  If we take seriously that octonions play some  underlying dynamical role
 in particle physics and SO(8) appears  as  a one-flavour unification group,
 then the  triality  property of SO(8) gives a natural reason for the
 existence of  three quark-lepton  generations.  Family   formation   from
  two flavours  due  to  weak  interactions  can  be  connected naturally
 enough to SO(10) group,  but  with  the  triality symmetry violated.
 An  attempt  to  restore  this  symmetry leads to the exceptional group
 $E_6$ and three quark-lepton families.

\end{document}